\documentclass[11pt,a4paper]{article}
\usepackage{amssymb}
\usepackage{amsmath}
\usepackage{graphics}
\usepackage{epsfig}
\usepackage{float}
\usepackage{caption2}
\usepackage{graphicx}

\begin{document}

\title{\textbf{\large{ Photodetachment of H$^{-}$ near a partial reflecting surface }}}
\vspace{1cm}
\author{A. Afaq\thanks{e mail: afaq@itp.ac.cn}\\
\textit{Institute of Theoretical Physics, Chinese Academy of
Sciences, Beijing 100080, China.}}
\date{}
\maketitle
\begin{abstract}
Theoretical and interpretative study on the subject of
photodetachment of H$^{-}$ near a partial reflecting surface is
presented, and the absorption effect of the surface is investigated
on the total and differential cross sections using a theoretical
imaging method. To understand the absorption effect, a reflection
parameter $K$ is introduced as a multiplicative factor to the
outgoing detached-electron wave of H$^-$ propagating toward the
wall. The reflection parameter measures, how much electron wave
would reflect from the surface; $K=0$ corresponds to no reflection
and $K=1$ corresponds to the total reflection.

 \vspace{.5cm}\noindent
\end{abstract}
\vspace{10pt} It has been observed both theoretically as well as
experimentally that the photodetachment cross section of H$^-$ shows
a smooth behavior in the free space \cite{Smith1959,Ohmura1960};
while in the presence of a reflecting surface it displays
oscillations \cite{YZC1,YZC2}. These oscillations are similar as if
they are in the presence of a static electric field
\cite{Fabrikant1980,Bryant1987,Stewart1988,Rau1988,Greene1988,Du1988,Kondratovich1990}.

Quite recently, Afaq and Du \cite{Afaq2007} have argued that this
oscillatory effect in the photodetachment cross section of H$^-$  is
because of two-path interference of the detached-electron wave from
the negative ion. To the observation point, one path comes directly
from the source H$^-$ , while the second path appears to be coming
from an image of the source behind the wall. In reference
\cite{Afaq2007}, the idea has been discussed without considering the
absorption effect of a wall. What would happen on the
photodetachment cross section of negative ion when the wall in use
will be absorbing? This problem is still interesting and has to be
discussed. I use a simple model for H$^-$ and provide quantitative
answer to the problem.

Near an absorbing wall the physical picture of the photodetachment
process may be described as: When the detached-electron wave is made
incident on the wall, a part of it is absorbed by the wall and the
other part is reflected with low intensity. This low intensity
electron wave propagates away from the system and appears to be
coming from an image behind the wall, and at a very large distance
it interferes with the direct outgoing detached-electron wave.
Consequently, we obtain an outgoing electron flux interference
pattern on the screen. The photodetachment cross section is
proportional to the integrated outgoing electron flux across a large
enclosure in which the source H$^-$ sits.

A partial reflecting wall $(0\leq K\leq1)$ is used for the electron
scattering and it is placed from H$^-$ at a distance more than $50$
Bohr radii, so that the asymptotic approximations can be valid. The
assumptions about the partial reflecting wall are the same as in
reference \cite{Afaq2007}. For an observer at large distance from
H$^{-}$, there are two components of detached-electron wave going
from H$^{-}$ to the observer. The first component propagates
directly from H$^{-}$ to the observer as if there is no wall; the
second component first propagates toward the wall, after being
partially reflected by the wall, it then propagates from the wall to
the observer. In the theoretical imaging method, the first component
comes from H$^{-}$ directly, the second component appears coming
from an image of the H$^{-}$ behind the wall. The detached-electron
flux can be calculated from the above two component outgoing waves.
By integrating the detached-electron flux for all angles, we are
able to derive analytic formula for the total photodetachment cross
section of H$^{-}$. Atomic units are used unless otherwise noted.

\begin{figure}
\epsfig{file=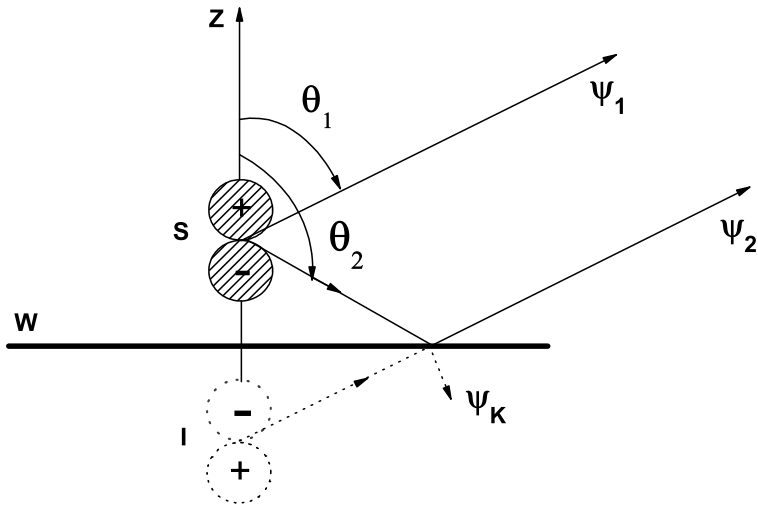,width=0.7\linewidth,clip=} \vspace{0.5cm}
\caption{\small The schematic diagram of the photodetachment of
$H^{-}$ near a wall. The negative ion $H^{-}$ is on the z axis, and
it's distance from the reflecting wall W is d. A z-polarized laser
light is applied for the photodetachment. The detached electron
waves at large distance include two components $\Psi_{1}$ and
$\Psi_{2}$. The direct component $\Psi_{1}$ is the detached electron
wave generated at the negative ion source S without the wall, the
second component $\Psi_{2}$ is obtained from $\Psi_{1}$ via a
reflection by the wall. The wave $\Psi_{2}$ appears to come from an
image I of the source S behind the wall W. $\Psi_{K}$ represents
absorbed component of detached-electron wave. The $\pm$ symbols
indicate the sign of two lobes of P-orbital wave function.}
\end{figure}

A schematic diagram for the photodetachment of H$^{-}$ near a
partial reflecting wall is shown in Fig. 1. A hydrogen negative ion
H$^{-}$ acting as a source (S) of detached electron wave is on
z-axis, its distance from the wall (W) is d. The reflecting surface
of the wall coincides with the x-y plane. A laser polarized in the z
direction is applied for the photodetachment of H$^{-}$. Three
components of detached electron wave $\Psi_{1}$,  $\Psi_{2}$ and
$\Psi_{K}$ are also shown. $\Psi_{1}$ is the direct component,
$\Psi_{2}$ is the reflected component and $\Psi_{K}$ is the absorbed
components by the wall. The reflected component appears as if it is
from an image (I) behind the wall. The $\pm$ symbols indicate the
sign of two lobes of P-orbital wave function.

The photodetachment process can be regarded as a two step
process\cite{Du1987,du1988,dU1988,Bracher}: in the first step, the
negative ion absorbs one photon energy $E_{ph}$ and generates an
outgoing electron wave; in the second step, this outgoing wave
propagates to large distances. Let $\Psi_{1}$ be the direct outgoing
electron wave after the photodetachment of H$^{-}$ in the absence of
the wall and let $\theta_1$ be the angle between the detached
electron and the z-axis. Half of this wave with $\theta_1$ smaller
than $\pi/2$ propagates away from the negative ion to large
distance. The other half of the wave with $\theta_1$ larger than
$\pi/2$ first propagates to the wall, after being partially
reflected by the wall, it then propagates to large distance. We call
this reflected wave $\Psi_{2}$. A part of this incoming wave towards
the wall is absorbed by the surface $\Psi_{K}$ which can be measured
by introducing reflection parameter $K$ as a multiplicative factor
to the wave moving toward the wall. The total outgoing electron wave
$\Psi^{+}$ at large distance from the system is given by
$\Psi^{+}=\Psi_{1}+\Psi_{2}.$

The expression for the direct wave $\Psi_{1}$ has been derived
before\cite{Du3}. Using $(r_1,\theta_1,\phi_1)$  as the spherical
coordinates of the electron with respect to the source (S) and
$(r_2,\theta_2,\phi_2)$ as the spherical coordinates of the electron
with respect to image (I), we have
\begin{eqnarray}
\Psi_{1}(r_{1},\theta_{1},\phi_{1}) &=&
U(k,\theta_{1},\phi_{1})\frac{\exp(ikr_{1})}{k r_{1}}, \nonumber \\
\Psi_{2}(r_{2},\pi-\theta_{2},\phi_{2}) &=&
U(k,\pi-\theta_{2},\phi_{2})K\frac{\exp i(kr_{2}-\mu\pi/2)}{k
r_{2}}.
\end{eqnarray}
Where $k=\sqrt{2E}$, and E is the detached-electron energy, $k_{b}$
is related to the binding energy $E_{b}$ of H$^-$ by
$E_{b}=\frac{k^2_{b}}{2}$, B is a normalization constant and is
equal to $0.31552$, and $K$ is the reflection parameter that
accounts how much electron wave reflects from the wall.
$U(k,\theta,\phi)$ is an angular factor, and for laser polarization
parallel to z-axis it can be written as
$U(k,\theta_{1},\phi_{1})=\frac{4k^2Bi}{(k_{b}^{2}+k^{2})^{2}}\cos\theta_{1}$,
~~~$U(k,\theta_{2},\phi_{2})=U(k,\pi-\theta_{2})=-\frac{4k^2Bi}{(k_{b}^{2}+k^{2})^{2}}\cos\theta_{2}.$

Eqs.(1) becomes
\begin{eqnarray}
\Psi_{1}(r_{1},\theta_{1},\phi_{1})&=&\frac{4k^2Bi}{(k_{b}^{2}+k^{2})^{2}}\cos(\theta_{1})\frac{\exp(ikr_{1})}{kr_{1}},\nonumber
\\ \Psi_{2}(r_{2},\theta_{2},\phi_{2})&=&-\frac{4k^2Bi}{(k_{b}^{2}+k^{2})^{2}}K\cos(\theta_{2})\frac{\exp
i(kr_{2}-\mu\pi/2)}{kr_{2}}.
\end{eqnarray}

Since $r_{1}$ and $r_{2}$ are large compared to the distance between
H$^-$ and the wall $d$, we can simplify further. Let
$(r,\theta,\phi)$ be the spherical coordinates of the
detached-electron relative to the origin. For phase terms, we
approximate as $ r_{1}\approx r-d\cos\theta$, $ r_{2}\approx
r+d\cos\theta$, and in all other places in Eqs. (2), we use
$r_{1}\approx r_{2}\approx r$, and $\theta_{1}\approx
\theta_{2}\approx \theta$. With these approximations, the outgoing
detached electron wave $\Psi^{+}$ from the system is given by
\begin{eqnarray}
\Psi^{+}(r,\theta,\phi)=\frac{4k^2Bi}{(k_{b}^{2}+k^{2})^{2}}
\cos(\theta) \left[e^{-ikd\cos\theta}-K e^{
i(kd\cos\theta-\mu\pi/2)}\right]\frac{\exp( i k r)}{k r}.
\end{eqnarray}
Eq. (3) represents the outgoing electron wave produced in the
detachment of H$^-$ near a partial reflecting wall. We now find
electron flux distribution on a screen at large distance and then
total photodetachment cross section. The electron flux is defined as
\cite{Afaq2007}
\begin{equation}
\vec{j}(r,\theta,\phi)=\frac{i}{2}(\Psi^{+}\vec{\bigtriangledown}\Psi^{+\ast}
-\Psi^{+\ast}\vec{\bigtriangledown}\Psi^{+}).
\end{equation}
Using the expression for $\Psi^{+}(r,\theta,\phi)$ in Eq. (3) and
flux formula in Eq. (4), we obtain the electron flux distribution
along the radial direction
\begin{equation}
{j_{r}}(r,\theta,\phi)=\frac{16k^3B^2}{(k_{b}^{2}+k^{2})^{4}}\cos^{2}(\theta)\left[\frac{1+K^{2}+2K\cos(2kd\cos\theta+\pi-\mu\pi/2)}{r^{2}}\right].
\end{equation}

We now calculate the total photodetachment cross section of negative
ion near a partial reflecting wall. Imagine a large surface $\Gamma$
such as the surface of a semi-sphere enclosing the source region, a
generalized differential cross section $\frac{d\sigma(q)}{ds}$ may
be defined on the surface from the electron flux crossing the
surface \cite{Afaq2007}, $\frac{d\sigma(q)}{ds}=\frac{2\pi
E_{ph}}{c}\vec{{j_{r}}}\cdot\hat{n}$, where c is the speed of light
approximately equal to 137 a. u., q is the coordinate on the surface
$\Gamma$, $\hat{n}$ is the exterior norm vector at q,
$ds=r^{2}\sin\theta d\theta d\phi$ is the differential area on the
spherical surface. The total cross section may then be obtained by
integrating the differential cross section over the surface,
$\sigma(q)=\int_{\Gamma}\frac{d\sigma(q)}{ds}ds.$ Therefore, the
first part of total photodetachment cross section of negative
hydrogen ion near a partial reflecting wall is
\begin{equation}
\sigma_1(E)=\frac{\sigma_{0}(E)}{2}\left[1+K^2-6K
A_1(2d\sqrt{2E})\right].
\end{equation}
with
\begin{equation*}
A_1(u)=\left[\frac{\sin(u-\mu\pi/2)}{u}+2\frac{\cos(u-\mu\pi/2)}{u^{2}}-2\frac{\sin(u-\mu\pi/2)}{u^{3}}-2\frac{\sin(\mu\pi/2)}{u^3}\right]
\end{equation*}

Where, $\sigma_{0}(E)=\frac{16\sqrt{2} \pi^{2}B^2
E^{3/2}}{3c(E_b+E)^3}$, is the photodetachment cross section of
$H^{-}$ in the absence of reflecting wall, the argument
$u=2d\sqrt{2E}$ of $A_1$  in Eq. (6) is equal to the action of the
detached-electron going from the negative ion to the partial
reflecting wall and back to the negative ion.

When the electron wave incidents on the surface of a wall, a part of
it is absorbed. Let we denote this part be $\Psi_K$ such that the
sum of the reflected part and the absorbed part would be equal to
the incoming electron wave to the wall. We introduce an absorption
parameter $T$ that measures how much of the electron wave is
absorbed by the surface of the wall such that $T^2+K^2=1$. The
absorbed part of the detached-electron wave is then given by
$\Psi_K(r,\theta,\phi)=T\Psi(r,\theta,\phi)$, where
$\Psi(r,\theta,\phi)$ is an electron wave from the source H$^{-}$
toward the wall. To calculate total cross section for the absorbed
part of the detached-electron wave, we performed similar steps as
for $\Psi^{+}(r,\theta,\phi)$ but integration limits for $\theta$
would be from $\pi/2$ to $\pi$. Hence, the second part of the total
cross section comes out
\begin{equation}
\sigma_2(E)=\sigma_{0}(E)\left[\frac{K^2-1}{2}\right].
\end{equation}
After adding the Eq. (6) and the Eq. (7), the total photodetachment
cross section for H$^{-}$  near a partial reflecting surface becomes

\begin{equation}
\sigma(E,K)=\sigma_{0}(E)A(2d\sqrt{2E})a^2_0.
\end{equation}
Where $A(u)$ is the modulation function and is defined by

\begin{equation}
A(u)=1-3K\left[\frac{\sin(u-\mu\pi/2)}{u}+2\frac{\cos(u-\mu\pi/2)}{u^{2}}-2\frac{\sin(u-\mu\pi/2)}{u^{3}}-2\frac{\sin(\mu\pi/2)}{u^3}\right].
\end{equation}

It is clear that Eq. (8) reduces to the case as there is no wall for
$K=0$. Hence for this particular condition, the wall acts like a
transparent medium for electron waves. For $K=1$  Eq. (8) reduces to
the results by Afaq and Du \cite{Afaq2007}.

\begin{figure}
\epsfig{file=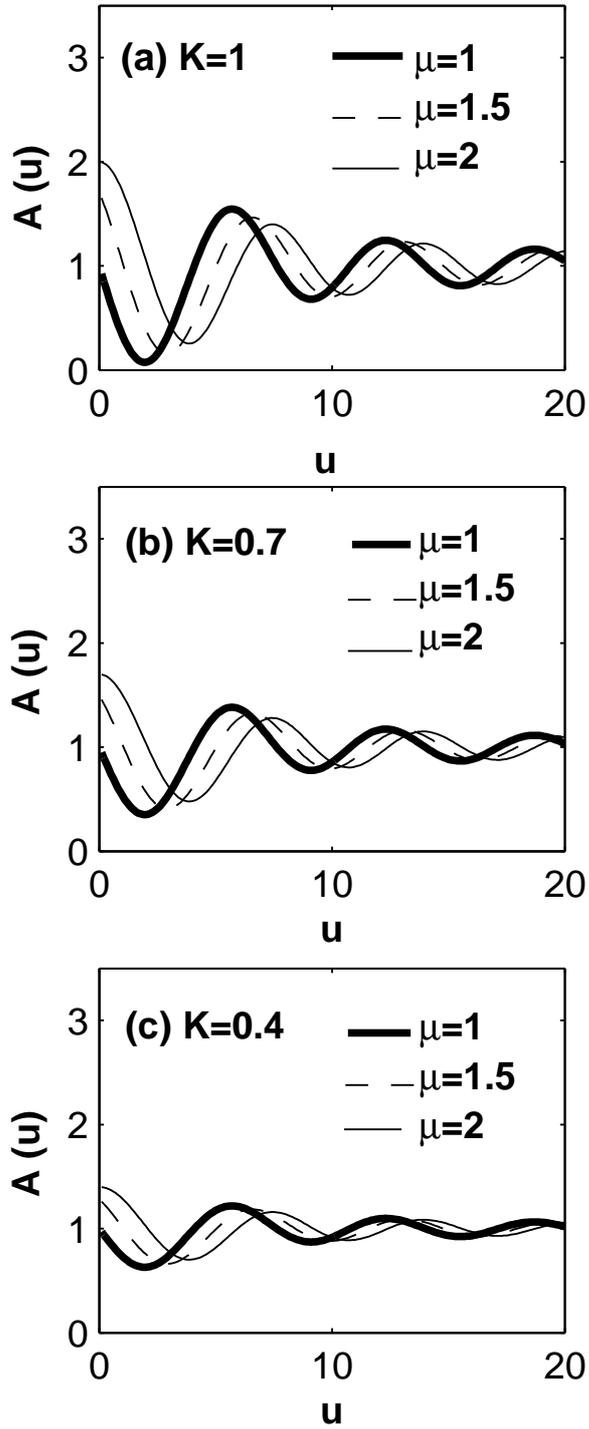,width=0.5\linewidth,clip=} \vspace{0.5cm}
\caption{\small The modulation function $A(x)$ is represented by
using Eq. (9) for different values of $K$ and $\mu$. One can observe
the change in oscillation amplitude and in phase with the change of
values of $K$ and of $\mu$ respectively.}
\end{figure}

\begin{figure}
\epsfig{file=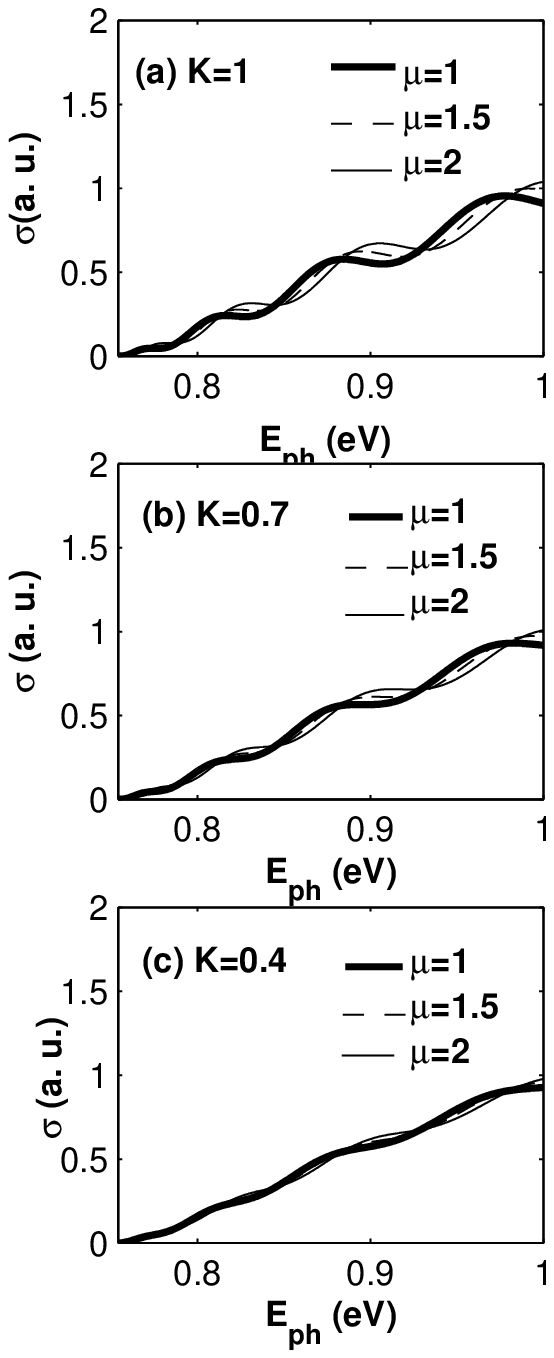,width=0.5\linewidth,clip=} \vspace{0.5cm}
\caption{\small The total photodetachment cross section  using Eq.
(8) with the modulation function in Eq. (9) for values of
$K=1,~0.7~,0.4$ and $\mu=1,~1.5,~2$. In these calculations, we used
$d=100$ Bohr radii.}
\end{figure}

Fig. 2 using Eq. (9) shows the behavior of the modulation function
$A(u)$ for different values of $K=1,~0.7,~0.4$ and $\mu=1,~1.5,~2$.
In Fig. 1(a), the soft wall case \cite{Afaq2007} is represented by
thick solid line and the hard wall case \cite{Afaq2007} is
represented by solid line, doted lines represent for the
intermediatory case. In Fig. 2 (a)-(c), we observe that the
amplitude of oscillation decreases and phase changes due to change
in values of $K$ and $\mu$. Fig. 3 using Eq. (8) with the exact
value of modulation function in Eq. (9) shows total photodetachment
cross section for the same values of $K$ and $\mu$ as in Fig. 2. For
$K=1$, the oscillation amplitude is large Fig. 2(a) and for $K=0.4$,
it becomes very small Fig. 2(c). It shows that oscillations in the
photodetachment cross section can effectively be controlled by
reflection parameter $K$.

The reason is, the electron wave that initially propagates toward
wall, a part of it is absorbed. Due to this absorption, the
reflected part will possess low intensity electron wave. This
reflected part appears coming from the image behind the wall. Two
path interference occurs on a screen placed at very large distance
from the system. Consequently, we observe a decrease in the
oscillation amplitude of photodetachment cross section.

Assuming a screen perpendicular to z-axis is placed at a distance L
from the wall, L is much greater than d and usually equal to
thousands atomic units in the experiments\cite{A11,A12}. The flux
distributions on the screen is cylindrically symmetric, it depends
only on the distance between any point $(x,y)$ on the screen and the
z-axis $\rho=\sqrt{x^2+y^2}$. By projecting the radial flux in Eq.
(5) along z-direction and then adding flux due to absorbed electron
wave function , the flux crossing the screen is
\begin{equation}
j_z(\rho)=\frac{32k^3B^2}{(k_{b}^{2}+k^{2})^{4}}\frac{L^{3}}{(\rho^{2}+L^{2})^{5/2}}\left[1+K\cos(\frac{2kdL}{\sqrt{\rho^{2}+L^{2}}}+\pi-\mu\pi/2)\right].
\end{equation}

\begin{figure}
\epsfig{file=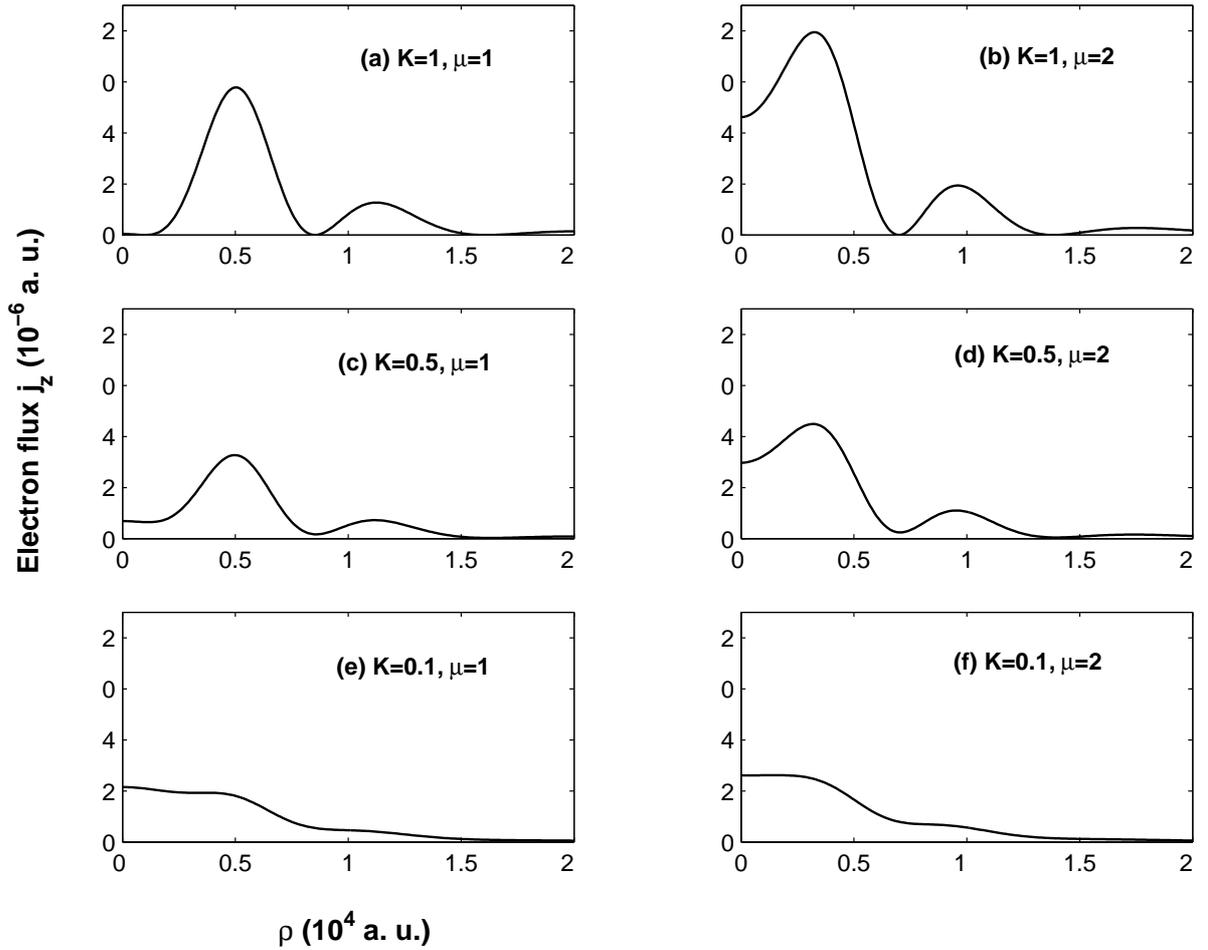,width=1.0\linewidth,clip=} \vspace{0.5cm}
\caption{\small The electron flux distributions using Eq. (10) on
the screen for values of $K=1,~0.5,~0.1$ and $\mu=1,~2$. We have
fixed photon energy $E_{ph}=1eV$ and the distance between the wall
and the screen $L=10000$ Bohr radii.}
\end{figure}


Fig. (4) using Eq. (10) represents the differential cross section
across z-axis for different values of $K=1,~0.5,~0.1$ and
$\mu=1,~2$. We have fixed photon energy $E_{ph}=1eV$ and the
distance between the wall and the screen $L=10000$ Bohr radii. Fig.
(4) represents the wall-induce interference for the differential
cross section, which may be observable in the photodetachment
microscopy experiments \cite{A11,A12}.

In summary, the photodetachment of $H^{-}$ near a partial reflecting
surface was studied using the theoretical imaging method. This
method made possible the derivation of analytical formulas of total
and differential cross sections in a straightforward manner. It is
concluded that there is a strong relation between the reflection
parameter $K$ and the oscillation of the cross sections. The
reflection parameter measures, how much of the electron wave would
reflect from the wall. For $K=0$  we get no oscillation, and for
$K=1$ we get maximum oscillation in the cross sections. On analyzing
the photodetachment spectra for different values of $K$, it may be
possible to characterize surfaces using an electron beam as a probe
just as various imaging and diffraction techniques, which have been
developed for surface study \cite{Yagi,Minoda}. The
detached-electron flux distributions on a screen placed at large
distance from the negative ion is also obtained. The distributions
displayed strong interference patterns. These patterns may be
observable just as in the photodetachment microscopy experiments for
negative ions in the presence of a static electric field
\cite{A11,A12,Du}. I hope that this theoretical study of
photodetachment of H$^-$ near a partial reflecting surface will
stimulate experiments, and may be useful in studying the surfaces.

I would like to thank Prof. M. L. Du for his useful discussion and
comments.

\end{document}